\documentclass{aastex61}

\received{}
\revised{}
\accepted{}
\submitjournal{AAS Journals}

\shorttitle{DECam: Size distribution of NEOs}
\shortauthors{Trilling et al.}
\begin{document}

\title{The size distribution of Near Earth Objects
larger than 10~meters}

\correspondingauthor{David E. Trilling}
\email{david.trilling@nau.edu}

\author{D. E. Trilling}
\altaffiliation{Visiting astronomer, Cerro Tololo Inter-American Observatory, National Optical Astronomy Observatory, which is operated by the Association of Universities for Research in Astronomy (AURA) under a cooperative agreement with the National Science Foundation.}
\affiliation{Department of Physics and Astronomy \\
P.O. Box 6010 \\
Northern Arizona University \\
Flagstaff, AZ 86011 }
\affiliation{South African Astronomical Observatory \\
PO Box 9 \\
7935 Observatory \\
South Africa}
\affiliation{University of the Western Cape \\
Bellville \\
Cape Town 7535 \\
South Africa}

\author{F. Valdes}
\affiliation{National Optical Astronomy Observatory \\
950 N. Cherry Avenue \\
Tucson, AZ 85719}

\author{L. Allen}
\altaffiliation{Visiting astronomer, Cerro Tololo Inter-American Observatory, National Optical Astronomy Observatory, which is operated by the Association of Universities for Research in Astronomy (AURA) under a cooperative agreement with the National Science Foundation.}
\affiliation{National Optical Astronomy Observatory \\
950 N. Cherry Avenue \\
Tucson, AZ 85719}

\author{D. James}
\altaffiliation{Visiting astronomer, Cerro Tololo Inter-American
  Observatory, National Optical Astronomy Observatory, which is
  operated by the Association of Universities for Research in
  Astronomy (AURA) under a cooperative agreement with the National
  Science Foundation.}
\affiliation{Cerro Tololo Inter-American Observatory \\
National Optical Astronomy Observatory \\
Casilla 603 \\
La Serena, Chile}

\author{C. Fuentes}
\altaffiliation{Visiting astronomer, Cerro Tololo Inter-American
  Observatory, National Optical Astronomy Observatory, which is
  operated by the Association of Universities for Research in
  Astronomy (AURA) under a cooperative agreement with the National
  Science Foundation.}
%\affiliation{Cerro Tololo Inter-American Observatory \\
%National Optical Astronomy Observatory \\
%Casilla 603 \\
%La Serena, Chile}
\affiliation{Departamento de Astronomia  \\
Universidad de Chile \\
Camino El Observatorio \#1515 \\
Casilla 36-D \\
Las Condes \\
Santiago \\ 
Chile}

\author{D. Herrera}
\altaffiliation{Visiting astronomer, Cerro Tololo Inter-American
  Observatory, National Optical Astronomy Observatory, which is
  operated by the Association of Universities for Research in
  Astronomy (AURA) under a cooperative agreement with the National
  Science Foundation.}
\affiliation{National Optical Astronomy Observatory \\
950 N. Cherry Avenue \\
Tucson, AZ 85719}

\author{T. Axelrod}
\altaffiliation{Visiting astronomer, Cerro Tololo Inter-American
  Observatory, National Optical Astronomy Observatory, which is
  operated by the Association of Universities for Research in
  Astronomy (AURA) under a cooperative agreement with the National
  Science Foundation.}
\affiliation{University of Arizona \\
Steward Observatory \\
933 N. Cherry Avenue \\
Tucson, AZ 85721}

\author{J. Rajagopal}
\altaffiliation{Visiting astronomer, Cerro Tololo Inter-American
  Observatory, National Optical Astronomy Observatory, which is
  operated by the Association of Universities for Research in
  Astronomy (AURA) under a cooperative agreement with the National
  Science Foundation.}
\affiliation{National Optical Astronomy Observatory \\
950 N. Cherry Avenue \\
Tucson, AZ 85719}



\begin{abstract}
We analyzed data from the first year of a
survey for Near Earth Objects (NEOs)
that we are carrying out with the
Dark Energy Camera (DECam) on the 
4-meter Blanco telescope at the Cerro Tololo
Inter-American Observatory.
We implanted synthetic NEOs into the data
stream to derive our nightly detection efficiency
as a function of magnitude and rate of
motion. Using these
measured efficiencies and the 
Solar System absolute magnitudes derived by
the Minor Planet Center
for
the 1377~measurements of 235~unique NEOs detected,
we directly derive, for the first time
from a single observational data set,
the NEO size distribution from 1~km
down to 10~meters.
We find that there are 
$10^{6.6}$~NEOs larger than 10~meters.
This result implies a factor of ten
fewer small NEOs than some previous 
results, though our derived 
size distribution is in
good agreement with several other estimates.
\end{abstract}




\section{Introduction}

Near Earth Objects (NEOs) are minor Solar System
bodies whose orbits bring them close to the Earth's
orbit. NEOs are important for both scientific investigations 
and planetary defense. Scientifically, NEOs, which
have short dynamical lifetimes in near-Earth space, 
act as dynamical and compositional tracers from elsewhere
in the Solar System. Studying NEOs also has the practical
application of searching for NEOs that could impact
the Earth and potentially cause widespread destruction.
Critically, the number of Chelyabinsk-sized bodies
(10--20~meters) is not well-constrained due to
various assumptions made in calculating
that population.
This leads to significant uncertainty in the impact risk of
these relatively common and relatively hazardous events.

Most NEOs are discovered by a small number of
dedicated surveys, including the Catalina Sky
Survey \citep{css},
the Pan-STARRS survey \citep{panstarrs},
and the restarted NEOWISE mission \citep{neowise}, and
more than 1000~NEOs are discovered every year.
The optical surveys (CSS, PS) use 1--2~meter class
telescopes, and their limiting magnitudes are roughly
V$\sim$21. The goal of these surveys is to discover
as many NEOs as possible, so any aspect of the survey
that diminishes discovery efficiency is eliminated.

We have carried out an NEO survey using
the 3~deg$^2$ Dark Energy Camera (DECam; \citealt{decam}) with the
4-meter Blanco telescope at the Cerro Tololo
Inter-American Observatory (CTIO);
for the purposes of moving object measurements,
this combination of camera and telescope
has been assigned the observatory code W84 by
the Minor Planet Center.
Our program was allocated 
10~dark nights per
``A'' semester for each of 2014, 2015, and 2016,
and 
is described in detail in Allen et al.\ (in prep.).
The etendue (product of aperture size and field of
view) of this survey is a factor of 2--10~larger
than that of other ground-based optical 
surveys, but our duty cycle of 10~nights per year
is quite small in comparison to the typical 200~nights
per year for dedicated surveys. The observational
niche of our new observing program is therefore not in discovering
a large number of NEOs, but rather (1) to discover
{\em faint} NEOs, through our much larger aperture, and
(2) to characterize our survey by implanting synthetic
objects in our data stream, allowing us to debias
and measure the size distribution of NEOs down to small
sizes. The large-scale dedicated surveys (CSS, PS) cannot
afford to detect and measure synthetic objects in
their data stream, which increases
processing time, but we can because of our comparatively short
observing season. 
Of course, no synthetic objects are
reported to the Minor Planet Center (MPC).

All NEO surveys are subject to observational
incompleteness that results in detecting a biased
sample. NEOs have a range of rates of motion, and
because the flyby geometries vary, optical brightness
does not necessarily correspond to NEO size.
In order to measure the underlying size distribution of NEOs,
knowledge of which NEOs are not detected is as important
as knowledge of which NEOs are detected. The best way to
measure this detection efficiency is through implanting
synthetic NEOs --- objects with the motions, PSFs,
noise properties, etc., of real NEOs --- into the data
stream, and then detecting the fake NEOs in the same 
way as real NEOs are detected. Thus, the detection 
efficiency of the survey
is readily measured as the number of synthetic
NEOs detected over the number implanted as a 
function of magnitude (or rate of motion or
orbital properties or any other aspect).
%No all-sky NEO survey implants synthetic
%NEOs in their data streams because they do
%not want to increase the data processing
%and validation aspects of their pipelines. 

Here we combine our detected (real) NEOs
with our measured detection efficiencies
to derive, for the first time, the debiased
size distribution of NEOs down to 10~meters
diameter as derived from a single
telescopic survey. We find a factor of ten fewer
10~meter-sized
NEOs than extrapolations from larger sizes
or normalization from terrestrial impact studies
predict. Some implications of this result
are discussed at the end of this paper.

\section{Observations and data processing \label{observations}}

%% In a manner similar to \objectname authors can provide links to dataset
%% hosted at participating data centers via the \dataset{} command.  The
%% second curly bracket argument is printed in the text while the first
%% parentheses argument serves as the valid data set identifier.  Large
%% lists of data set are best provided in a table (see Table 3 for an example).
%% Valid data set identifiers should be obtained from the data center that
%% is currently hosting the data.

A detailed description of the observing
cadence, sky coverage, filters, and exposure
time is presented in Allen et al.\ (in prep.).
Here we use only results from our ten-night observing run
in April/May, 2014.
Briefly, each survey field (3~deg$^2$) was typically
observed 5~times per night and on 3~nights.
We observed using the broad VR filter.

The data processing steps are also presented
in detail in Allen et al. 
In summary, each exposure is flat fielded and astrometrically calibrated using the standard NOAO Community Pipeline (CP) for DECam \citep{valdes}.  A photometric zeropoint, which leads to the reported magnitudes, is computed by matching stars to Pan-STARRS magnitudes \citep{magnier}.
However, in fields for which Pan-STARRS magnitudes were not available, the reference catalog used is the USNO-B1 photographic catalog \citep{usno} with a transformation designed to match, on average, the more accurate CCD magnitudes.
For fields for which Pan-STARRS photometry is available
we transform the catalog g and r to V using a transformation from Lupton (2005) and then match the observed VR instrumental magnitudes to those values.
For non-Pan-STARRS fields we
use USNO-B1 photographic magnitudes transformed to r \citep{usno},
which gives us the pseudo-r magnitude.
For the purposes of this paper, we treat all magnitudes
that we reported to the MPC as~V. We estimate
that the photometric errors are typically less than~0.1~magnitudes,
but formally use 0.2~mag here to include errors introduced
by these various transformations.
Our detection limit is around
SNR$\approx$5 for objects that are not trailed
or only slightly trailed;
for trailed objects,
our detection limit corresponds to SNR$\approx$5~per pixel
at
the brightest part of the trail.

A special version of the CP incorporating the NOAO Moving Object Detection System \citep{mods}
adds the following steps.  Exposures from each survey field in a night are median-combined to provide a reference image with transients removed. The median is subtracted from each contributing exposure to remove the static field. Catalogs of transient sources are created from the difference image.
Objects with
3~or more detections with similar magnitudes that make a track of a consistent rate and with shapes (elongation/P.A.) consistent with that rate are identified as
candidate moving objects.
All objects with
{\tt digest2} scores\footnote{{\tt https://bitbucket.org/mpcdev/digest2}}
greater than~40\% --- that is, where the probability of being
an NEO or other interesting object (Trojan, etc.),
based on the short orbital arc, is greater than 40\% ---
are verified through visual inspection to
eliminate false detections (chance cosmic
ray alignments, etc.). 
All validated objects are submitted to the 
MPC; this list includes NEOs as well
as 
many valid
moving objects 
that are not NEOs.

\begin{figure}
\includegraphics[angle=270,scale=0.6]{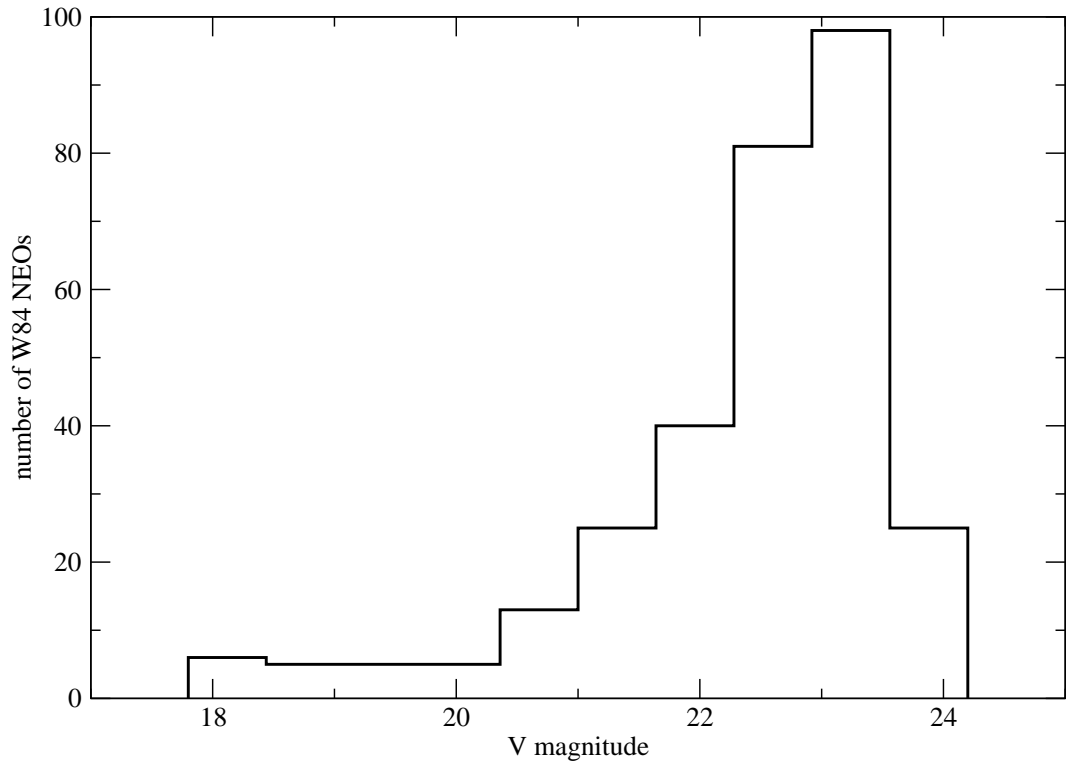}
\caption{Histogram of transformed V magnitudes for
all NEOs with preliminary or permanent designations
that were
detected in our DECam NEO survey (1377~observations
of 235~objects).
We show the first reported V magnitude for each object.
\label{vmag}}
\end{figure}

Postage stamps of several representative NEOs
observed by us are shown in Figure~\ref{stamps}.
A histogram of V magnitudes of our detected
(real) NEOs is shown in Figure~\ref{vmag}.

\section{Detection efficiency}

\begin{figure}
\includegraphics[scale=0.25]{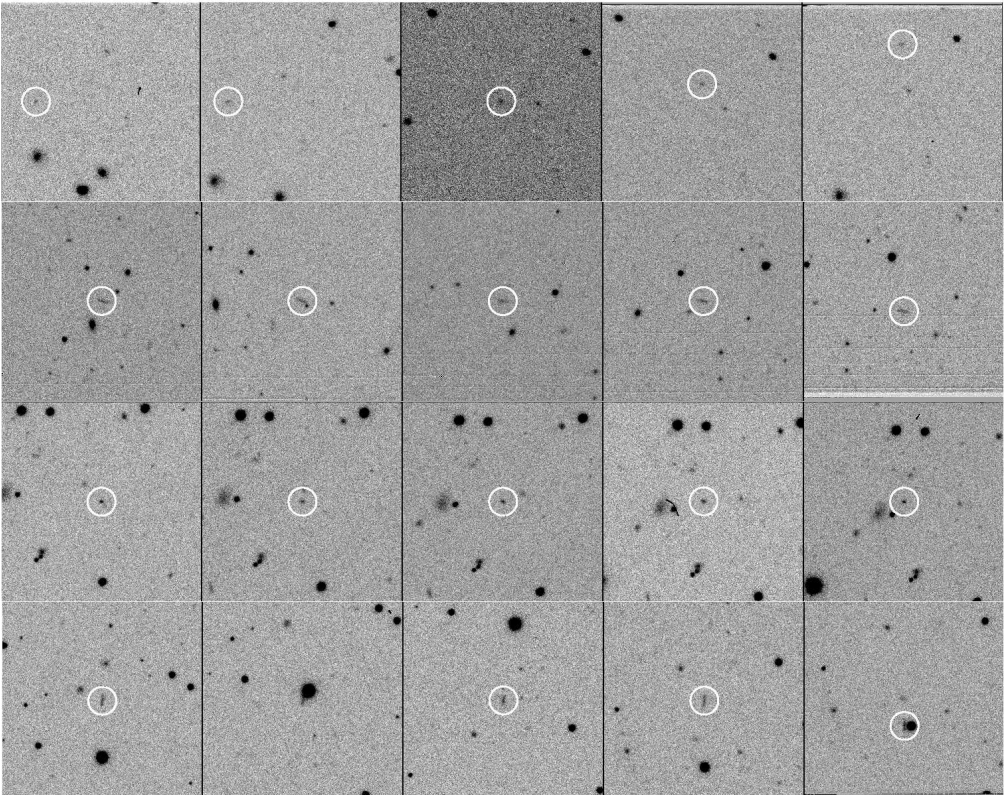}
\caption{Postage stamps showing, from top to bottom, detections of 
real objects AiK41PR (rate of motion 138"/hr, magnitude 23.0) and
AiNc1M6 (407"/hr, 22.3)
and synthetic objects
Sim\_AiNd1U6 (98"/hr, 22.3), and
Sim\_AiNd1O1 (334"/hr, 21.3).
From left to right in each row are the
five images in our detection sequence,
with typical time from the first to last
image of around 20~minutes.
\label{stamps}}
\end{figure}

\begin{figure}
\includegraphics[angle=270,scale=0.6]{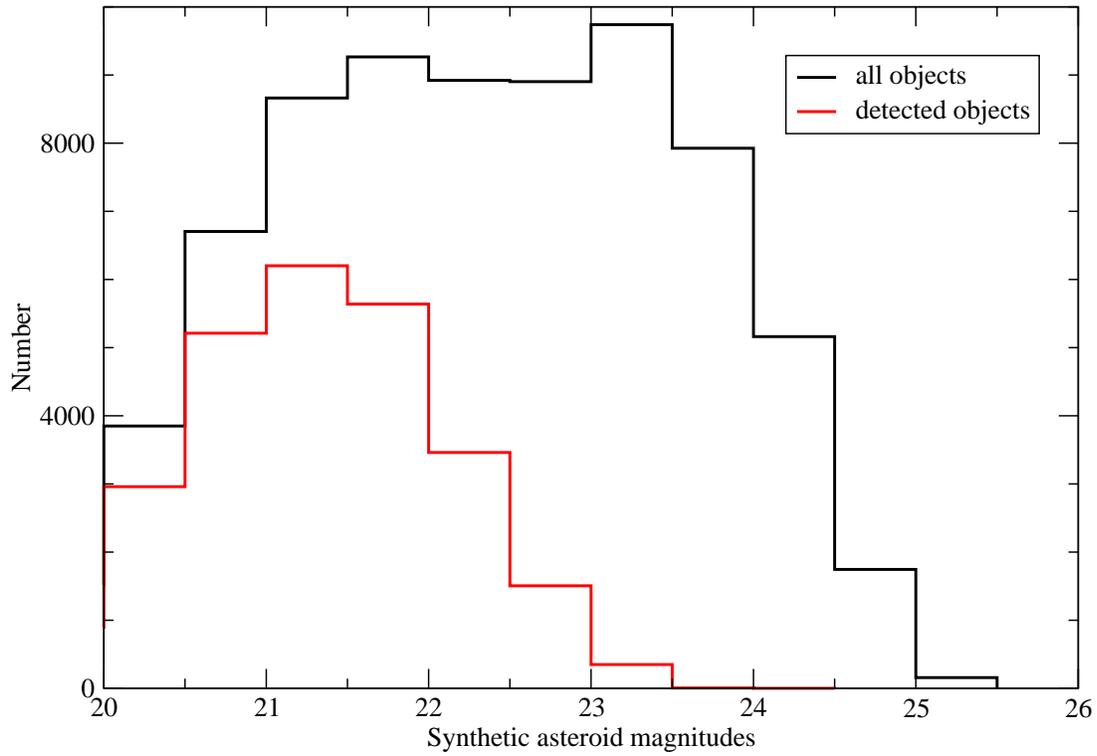}
\caption{Magnitude distribution of synthetic
(implanted) objects. The implanted population is not
designed to mimic any particular underlying
NEO population, but the overall shape is
roughly similar to the expected distribution
for a complete survey.
The numbers shown on the vertical axis
are in units of nightly tracklets, each
of which consists in most cases
of five observations, so, as an example,
more than 50,000~individual NEO
point sources were implanted with 
magnitudes in the range 25--25.5.
\label{magdistrib}}
\end{figure}

\begin{figure}
\includegraphics[angle=270,scale=0.6]{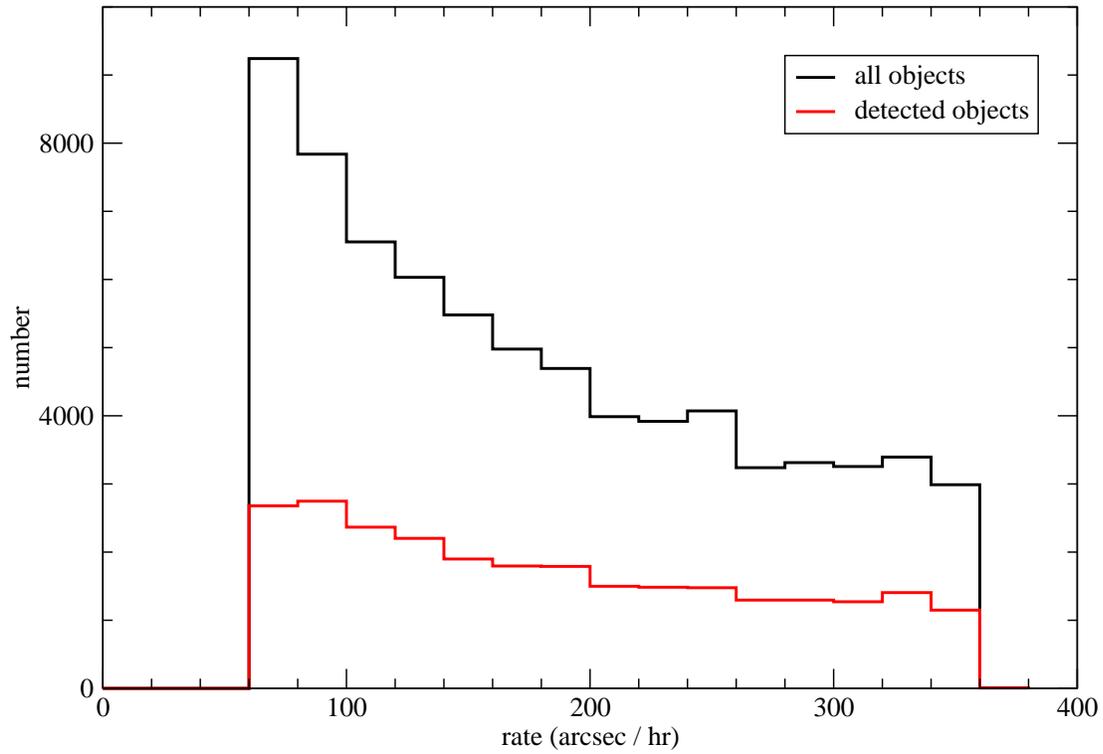}
\caption{Rate distribution of synthetic
(implanted) objects.
The implanted population is not
designed to mimic any particular underlying
NEO population, but the overall shape is
roughly similar to the expected distribution
for a complete survey.
We detected a handful of real NEOs with rates
greater than the fastest synthetic NEO rate
of 360~arcsec/hour.
As in Figure~\ref{magdistrib}, the numbers
on the vertical axis are tracklets, not point source
images.
\label{ratedistrib}}
\end{figure}

One hundred synthetic NEOs, i.e., fake asteroids, 
are created in a square that circumscribes each
pointing; on average, around 72~objects fall within
the DECam field of view and not in gaps.
These synthetic objects are 
implanted directly in each exposure. 
This is many more synthetic NEOs than real NEOs in each exposure,
and allows us to probe the details of our sensitivity with
a large number of objects.
Over the entire observing run, around 
365,000~synthetic asteroid detections were
possible.
The distributions of magnitude and rate of motion
for the synthetic population are not matched
to any specific underlying NEO population but
do generally approximate an observed NEO
population, while extending to much fainter
magnitudes than could be detected in this survey
(Figures~\ref{magdistrib} and~\ref{ratedistrib}).
The synthetic objects are created independently for each field with simple linear motions; there
 is no linkage across fields or nights. 
The detected synthetic objects are identified based on their known
implant positions. We recovered the synthetic asteroids in the same way as real asteroids, that is, satisfying the same minimum number (3) of observations per field, up to the point of 
running {\tt digest2} (all synthetic objects would have
high NEO probabilities), visual inspection (since these
are known to be valid objects), and, naturally, reporting
to the MPC. Postage stamp images of representative
synthetic images are shown in Figure~\ref{stamps}.

Two important features of the synthetic implants for the efficiency characterization are the seeing and streaking.  The seeing of each exposure was used to provide a point-spread-function (PSF) and the static magnitude of the source was trailed across the image based on its rate (as shown in Figure~\ref{stamps}).  These aspects affect the surface brightness, which means that the detection efficiency varies with the conditions on each night and field and the apparent rate of motion.

\begin{figure}
\includegraphics[angle=270,scale=0.6]{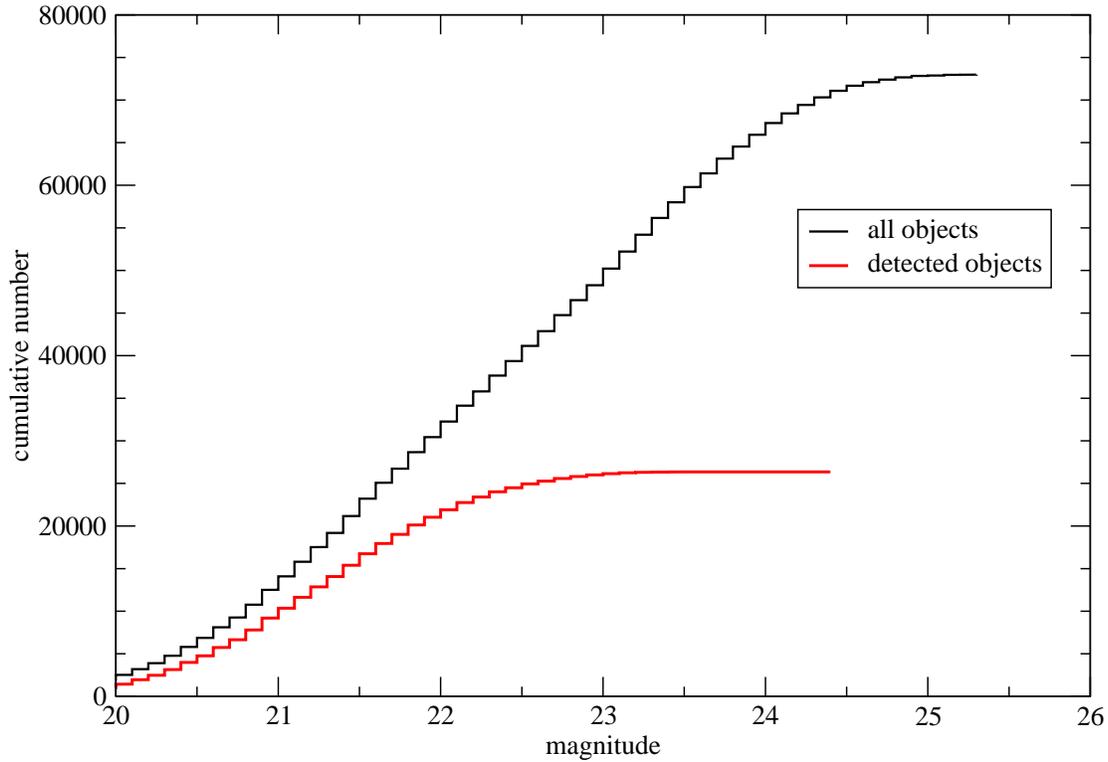}
\caption{Cumulative number of implanted (dashed line)
and detected (solid line) synthetic NEOs brighter
than a given magnitude. By design, objects are implanted
beyond our expected detection limit so that our detection
efficiency can be measured well at the faint end. Few
synthetic objects fainter than~24 were detected,
as shown by the flattening of the solid line.
As in Figure~\ref{magdistrib}, the numbers
on the vertical axis are tracklets, not point source
images.
The ``stair-step'' pattern here simply indicates
that our synthetic objects have magnitudes
given only to 0.1~mag.
\label{synthetic}}
\end{figure}

We note that our debiasing procedure requires
the reasonable assumption that NEO sizes
and albedos (in other words, their absolute
magnitudes) are indepedent of flyby geometry
(distance from the Earth, phase angle, etc.).
Debiasing must only take into account     
the survey properties that bias our observed
sample: magnitude and rate of motion, but not
geometry.
The measured detection efficiency as a function
of magnitude is easily calculated as the 
number of synthetic NEOs detected divided by
the number of synthentic NEOs implanted.
The detection efficiency is calculated
for each night and for each of four bins in
rate of motion (60--135, 135--210, 
210--285, 285--360~arcsec/hr).
Our measured detection efficiencies as a function
of observed magnitude are shown
in Figure~\ref{deteff}.
The overall detection efficiency for all synthetic
objects as a function of H~magnitude is
shown in Figure~\ref{synthetic}.

For real objects, there are 303~unique ``object-nights'': a given
object observed on a given night. As an example, 
2014~HA$_{196}$ was discovered by us on 20140422
and re-observed by us on 20140427. This asteroid
therefore
has two ``object-nights'' and appears twice
in our list of 303~``object-nights,'' giving
us two different opportunities to debias the 
NEO population with this asteroid. Because we 
normalize our resulting size distribution (see below),
counting individual objects more than once does
not introduce a significant error for our result.

\section{The size distribution of NEOs}

Observations that meet the following criteria
are used in this debiasing work:
(1) the object observed must be classified as
an NEO by the MPC;
(2) the object observed has received either
a
preliminary designation (such as
2014~HD196, one of the W84
discoveries from the 2014 observing season) or permanent
designation (for example, asteroid 88254,
for which our nine W84
observations over two nights are only a small fraction of
the more than 400~observations of this asteroid to date),
which means that the orbit is relatively well
known and therefore that both its NEO status
and H~magnitude are reasonably secure;
and
(3) the observations were made by us, i.e.,
observatory code W84.
There are a total of
1377~measurements of 235~unique objects
that meet these criteria.
97~of these objects were discovered
by our survey.

\begin{figure}
\includegraphics[angle=270,scale=0.6]{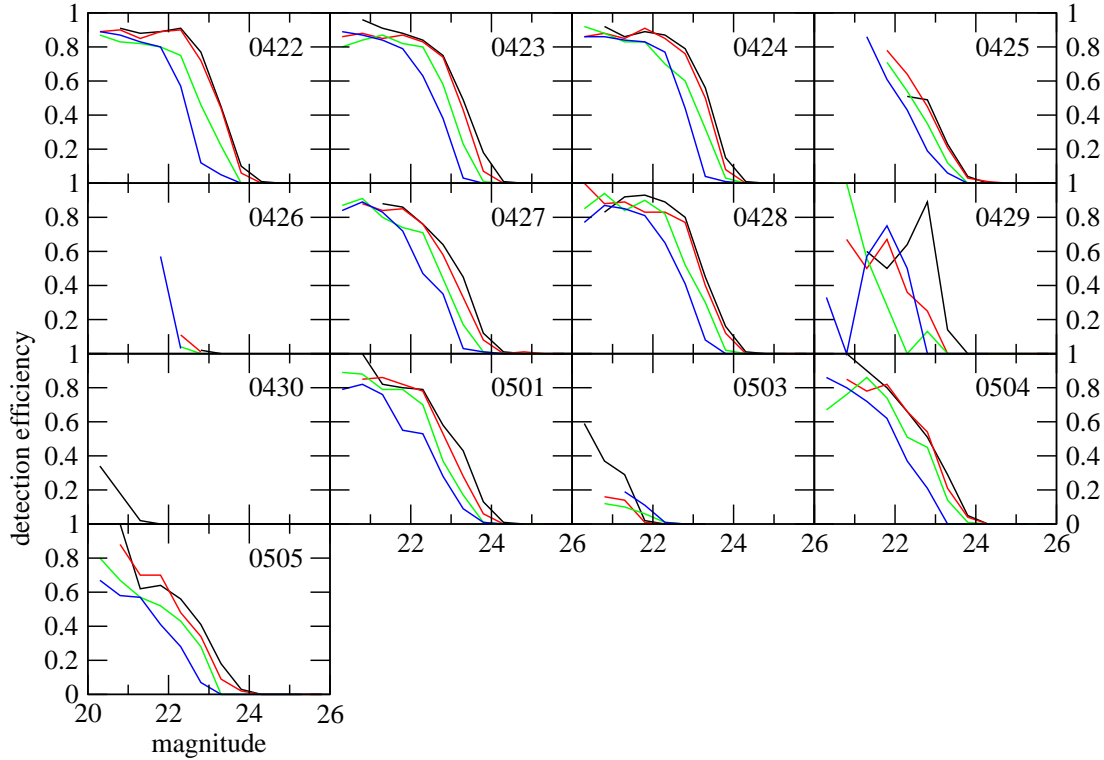}
\caption{Nightly detection efficiencies as a function
of magnitude and rate of motion.
For all panels, the colors are as follows:
black (60--135~arcsec/hr);
red (135--210~arcsec/hr);
green (210--285~arcsec/hr); and
blue (285--360~arcsec/hr).
There are $\sim$72~synthetic objects inplanted
within each image, resulting in more 
than 10,000~synthetic objects in each night.
The UT date at the beginning of the night is given
as MMDD. Nights with poor conditions and poor
transparency are evident (0430, 0503).
Night 0426 was our first experiment with
implanting synthetic objects (which was not carried
out on the nights in chronological order) and has fewer
measured than other nights, leading to
poorly defined efficiency curves.
The efficiencies shown here are nightly averages.
For most nights, the maximum efficiency, for 
slow and bright objects, is around 0.8--0.9.
Typical 50\% efficiency is around magnitude~23,
though this depends on rate of
motion and conditions on a given night;
our ability to detect fast objects
is not as good as our ability to detect
slower objects.
In our debiasing each object is debiased 
according to the efficiency for the night
of the observation in question and the 
magnitude and rate of motion of the object.
\label{deteff}}
\end{figure}

When detected NEOs and the detection 
efficiency are both known, deriving the
size distribution of NEOs is straightforward.
Each $i$th NEO is detected at magnitude
$V_i$ and rate of motion $r_i$. 
The detection efficiency at 
that magnitude $\eta_i(V_i,r_i)$ is known (Figure~\ref{deteff}).
Each $i$th detected asteroid therefore, when
debiased, represents 
$N_i = 1/ \eta_i$ asteroids,
applying the correction
for the number of NEOs of that magnitude and
rate of motion
that exist but were not detected in our 
survey. 
Note that each NEO is debiased
individually using the appropriate $\eta$ for
that object, using the nightly efficiencies
shown in Figure~\ref{deteff}.

\begin{figure}
\includegraphics[angle=270,scale=0.6]{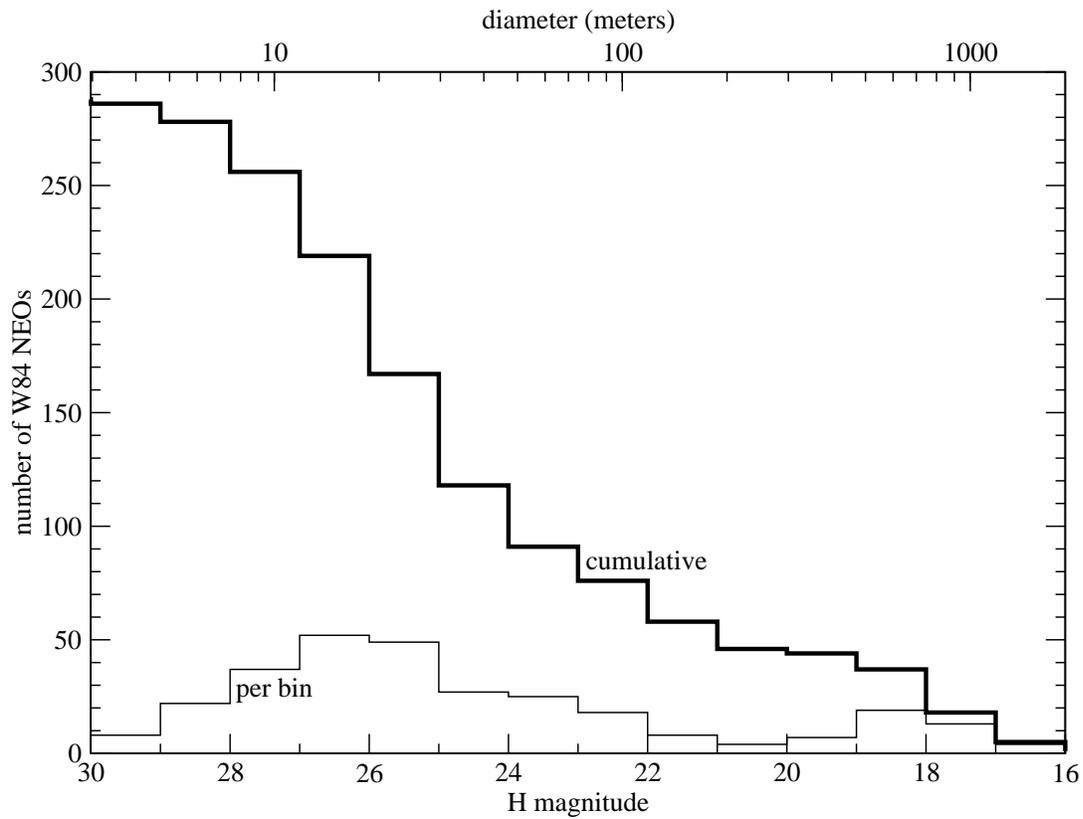}
\caption{Histograms of H magnitudes (Solar System
absolute magnitudes) [bottom axis]
for all NEOs with preliminary
or permanent designations that were observed in our
DECam NEO survey. The top axis shows diameters,
assuming that each object has an albedo of~0.2. 
The thin line shows the number of objects
in each bin (0.5~mag wide), and the thick line
shows the cumulative number brighter than a given
$H$.
\label{hmag}}
\end{figure}

The observed V magnitudes do not 
specify the size of the asteroid.
The MPC processes
the submitted astrometry to derive
an orbit. Given that orbit and the reported
magnitude, the Solar System absolute
magnitude $H$ (the hypothetical magnitude
an object would have at 1~au from the 
Sun, 1~au from the observer, and at zero
phase)
can be derived.
For each $i$th asteroid we use the 
MPC-derived absolute magnitude $H_i$.
Figure~\ref{hmag} shows the histogram
of all $H$ magnitudes in our survey.
Each $i$th asteriod therefore represents
$N_i$ asteroids with absolute magnitude $H_i$.
Finally, we derive the cumulative size distribution
$N(<$$H)$
by summing all $N_i$ for a given $H\leq H_i$.

An asteroid's diameter $D$ and $H$ are related through
the albedo $p_V$ as

\begin{equation}
D = \frac{1329}{\sqrt{p_V}} 10^{-H/5}
\end{equation}

\noindent \citep{chillemi}. We know nothing about the
albedos of the NEOs we observed. Therefore,
to calculate asteroid diameters 
we
adopt an albedo 
of~0.2, which is the average albedo
from \citet{mainzer} for bodies 
smaller than 200~meters.
We adopt this NEOWISE albedo value since their
survey is relatively insensitive to asteroid albedo,
unlike other NEO surveys;
we discuss the uncertainties introduced
by this assumption of a single average
albedo below.
There is therefore a direct translation
from the absolute magnitude distribution
to NEO size distribution.
We calculate the cumulative size distribution:
the number of objects larger than
a given diameter.

The final step in deriving the size distribution
of NEOs is a correction for the volume searched.
This is analogous to the Malmquist 
bias present in flux-limited astrophysics
surveys\footnote{This is similar in concept
to the kind of analysis described in
\citet{schmidt}, although the details
of the corrective approach differ.}:
we can detect 100~meter
NEOs at a greater distance (and therefore in 
a greater volume) than 
we can detect 10~meter NEOs.
The NEO diameter $D$ is given by

\begin{equation}
D~{\rm (m)} = 2\times \sqrt{ \left(\frac{f_{NEO}}{f_\odot}\right)
\left(\frac{\Delta^2}{p}\right)\left(\frac{R}{1.5\times10^{11}~{\rm m}}\right)^2}
\end{equation}

\noindent where
$f_{NEO}$ and $f_\odot$ are the fluxes from the NEO
and the Sun, respectively;
$\Delta$
and $R$
are the geocentric and heliocentric
distances of the NEO
at the time of the flux measurement, in 
meters; and $p$ is the
albedo of the NEO.
For opposition surveys such as ours, 
$R$ can be approximated as
$\Delta + 1.5\times 10^{11}$~meters, so to 
first order
$R^2 \Delta^2$ is approximately $\Delta^4$.
For a given flux limit $f_{NEO}$ and constant
albedo $p$, diameter is therefore
proportional to $\Delta^2$.
The ratio of the (geocentric) search distances for any two
sizes $D_a$ and $D_b$ is therefore 
$\sqrt{D_a/D_b}$, and the ratio of 
searched volumes is $(D_a/D_b)^{1.5}$.

Most of our objects were detected with 
$\Delta$$<$1~au, and all of them with
$\Delta$$<$1.3~au.
Our detection limit of V$\sim$23
for bodies at 1.3~au corresponds to 
NEOs with sizes $\sim$200~meters.
Thus, our survey is complete for 
objects larger than 200~meters ---
that is, we would have detected every NEO
larger than 200~meters that appeared within
our search cone ---
and
we need only apply the volume correction
described above for objects smaller
than 200~meters (around H=21).
Therefore,
we set 
$D_a$ above to be 200~m.
We 
multiply our measured and debiased size distribution
at every $D_b$ smaller than 200~m by
the factor $(200~{\rm m}/D_b)^{1.5}$ to correct
for this volume effect. This has the effect
of ``adding back'' small ($<$200~m)
NEOs that would have been in our search
cone but too faint to have been detected
(through being too distant).

The choice of 1.3 au as the outermost boundary
of NEO space is well justified.
Around 50\% of our discovered NEOs have geocentric distances between 1.0~and 1.3~au,
and none beyond 1.3~au. 
Our discovered
distribution of some 70 objects indicates that the detection volume is well-sampled
and extends to 1.3~au.

Finally,
our survey was relatively small, covering
$\sim$975~square degrees in 348~distinct
pointings. We therefore normalize
our derived size distribution to results
from NEOWISE and ExploreNEOs, both of which
independently derive the result that
there are around 1000~NEOs larger than
1~km \citep{mainzer,trilling}, in agreement with an earlier estimate
by \citet{werner}.

Our final result 
is shown in Figure~\ref{sizedist}.
We find around $10^{6.6}$~(or $3.5\times 10^6$) NEOs larger than
H=27.3 (around
10~meters), and 
$10^{6.9}$~(or $7.9\times 10^6$)~NEOs larger than H=28 (around
7~meters).
For the first time, the size
distribution of NEOs 
from 1~km to 10~meters has 
been derived from a single dataset.
This is significant as it bridges previous
observational results that had data only at either
the 
large or small end of this range,
as discussed below.

\section{Discussion}

\subsection{Objects not included in this analysis}

We include in this analysis only objects that
have been designated and identified as
NEOs by the MPC.
This could introduce biases in two ways,
as follows.

First, we report to the MPC only objects
that have {\tt digest2} scores of $>$40\%.
We have not yet reported the many thousands
of objects that we detected that have low {\tt digest2}
scores. These are objects that are
likely main belt asteroids and have 
low probabilities of being an NEO.
However, even if an object
with a low {\tt digest2} score
is an NEO,
the low probability indicates that it is
probably moving slower than typical
NEO rates, which means that it is relatively far
from the Sun and Earth. In this case, the object
would have to be relatively large to be bright
enough to be detected by us, and would therefore
have little effect on the derived size distribution of 
small NEOs.

In order to further understand any potential
bias introduced by using {\tt digest2} we carried
out the following experiment.
We chose ten random recently discovered objects from the
MPC's list of NEOs, and for each
ran only the first five measurements from the 
first night of observations through {\tt digest2}.
For these ten objects, the lowest {\tt digest2}
score was~78, and seven of the objects had
{\tt digest2} NEO scores of~100 (i.e., 100\% probability
of being NEOs).
We also chose five measurements from the discovery
nights for ten random Hungaria asteroids and ten
random Mars crossing asteroids.
The Hungarias have {\tt digest2} scores in the
range~17--66, and the Mars crossers have scores~14--91.
From this experiment we conclude that 
non-NEOs can sometime be misclassified as NEOs
on the basis of their {\tt digest2} scores (given
our $\geq$40\% threshold), but
no NEOs are ever misclassified as non-NEOs.
We therefore do not miss any NEOs through this
{\tt digest2} filtering. 
Since the only objects
used in this analysis are those with
preliminary designations --- those with orbits
that are conclusively NEOs --- our approach is
unlikely to have either false negatives (missing
objects that should be included) or 
false positives (including objects that should
not be).

\begin{figure}
\includegraphics[angle=270,scale=0.6]{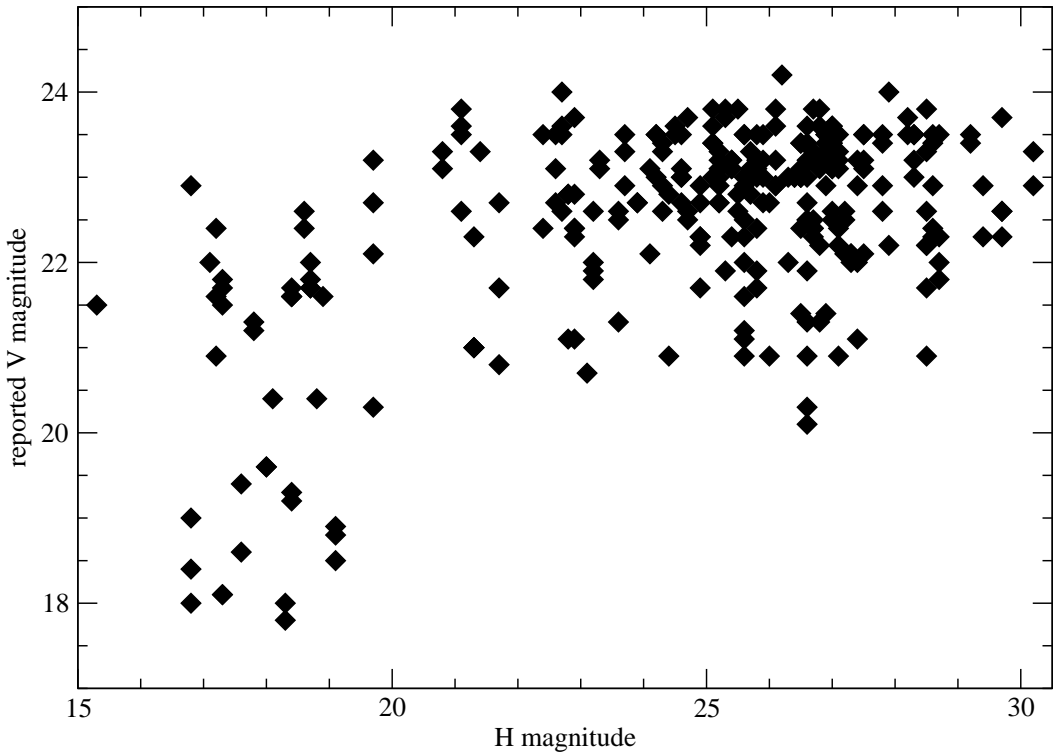}
\caption{Reported V magnitude as a function
of H magnitude for the objects used to 
calculate the NEO size distribution. For
V$>$20 and H$>$20 there is no correlation
between V and H, so a bias against recoveries
and designations of the faintest (in V) objects
does not affect the derivation of the underlying
size distribution (that is, H magnitude 
distribution).
\label{HV}}
\end{figure}

Second,
objects that we observed but that never were
designated are probably the faintest objects,
because neither we nor any other facilities were
able to recover them. 
%This could reasonably 
%apply for objects with V$>$20. 
Figure~\ref{HV} shows that 
there is no correlation between H and V for V$>$20
(a conservative limit for recovery facilities,
which generally use telescopes in the 
1--2~meter class, though some smaller telescopes
also contribute)
and H$>$20
(in other words, smaller than a few hundred
meters diameter).
This lack of relationship between H and V 
is because there is no dependence of orbital elements
on asteroid size.
Therefore, there is no bias introduced in our 
small NEO
size distribution
calculation even though
targets with the faintest (apparent) magnitudes
may
have been
preferentially omitted through a recovery
bias. Observed magnitude (and therefore
recovery probability) is essentially independent
of size for NEOs smaller than around 300~meters.

%--by normalizing at 1 km we are also normalizing at H=20, which is where the non-correlation of V-H kicks in

\begin{figure}
\includegraphics[angle=270,scale=0.6]{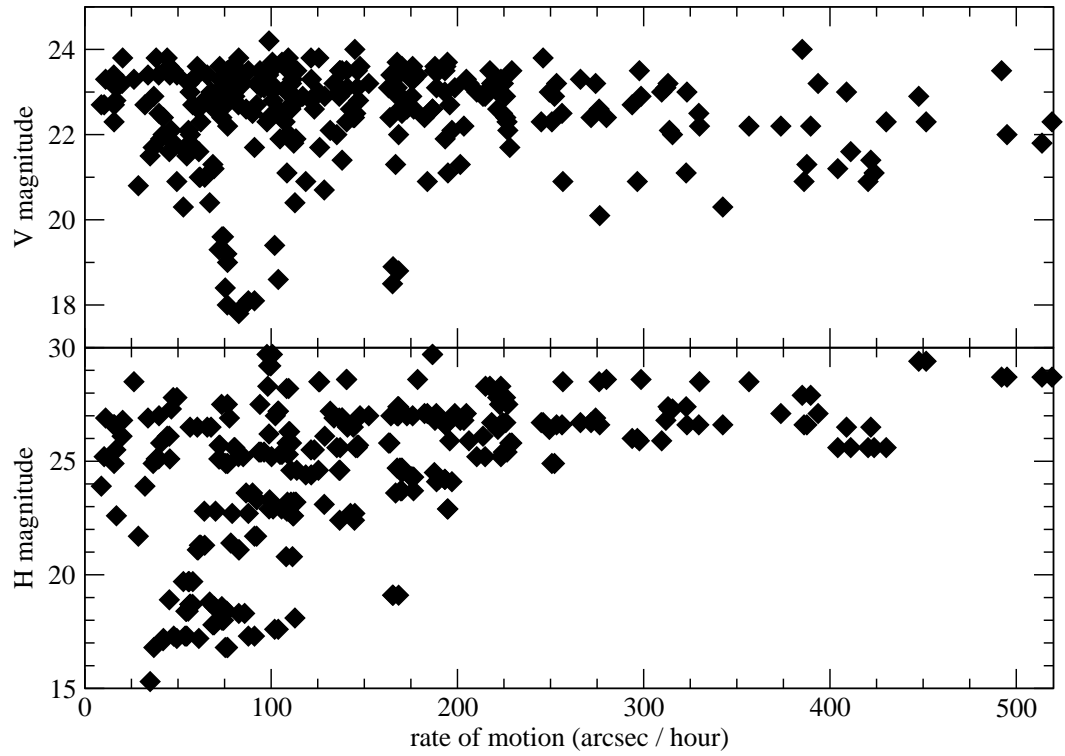}
\caption{Apparent V magnitude and 
H magnitude as a function of 
rate of motion for
all W84 objects reported here.
Some 6.5\% of objects have
rates faster than 360~arcsec/hour;
for H$>$25 (bodies smaller than 30~meters),
the fraction is around 10\%.
We therefore
estimate roughly that the number of small
(H$>$25) NEOs that we find could
be underestimated by some 10\%.
\label{rateofmotion}}
\end{figure}

Finally, there is a chance that some of the fastest
moving objects were missed by our automated detection
algorithms. Using this set of synthetic objects,
we cannot calibrate our detection efficiency
for (real) objects moving faster than 360~arcsec/hr because 
no synthetic objects moving faster than this rate were
implanted in the data stream (Figure~\ref{ratedistrib}).
For objects moving faster than this rate,
we extrapolate our derived efficiencies
(as a function of magnitude, and for the 
appropriate night)
for the fast-moving object in question, and
use these extrapolated efficiencies in our
cumulative size distribution calculation.
The effect of this extrapolation is small.
Figure~\ref{rateofmotion} shows, for our real
objects, the measured rate of motion and the
V and H magnitudes of those detected objects.
We find that some 6.5\% of real objects have
rates faster than 360~arcsec/hour; for
H$>$25 (bodies smaller than 30~meters), the fraction is around 10\%.
So, while the extrapolated efficiencies are only
an approximation,
the relatively small uncertainty introduced affects only a small
number of objects.

In conclusion, omitting some objects from this analysis
for the reasons explained above likely does not introduce
a significant error to our derived NEO size distribution,
though we may underestimate the number of 
small NEOs by some 10\%.
We therefore conclude that the sample we consider
here can be used to derive the small NEO size distribution
without any large bias from reporting or
recoveries.

\subsection{Uncertainties}

%xxx
%So, just as we do for efficiency, we could use the implanted source information to estimate total light (as a function of mag/rate).  I'm not keen on undertaking that but it at least needs to be recognized that the H mags are based on what observers report (and it is not just us) and these are likely to be biased by missed light for non-point (moving streak) sources.
%xxx

Several uncertainties remain in the size distribution
estimate shown in Figure~\ref{sizedist}. One is that the
photometry of our survey has uncertainties on the order
of 0.1~mag, though, as described in Section~\ref{observations},
we formally assume photometric uncertainties of 0.2~mag.
These V~magnitude uncertainties transfer directly
to H~magnitude uncertainties\footnote{
It is important to note 
that in this study we use only designated NEOs, for
which orbital information is known; if we had used
objects without good orbits then the calculation of
H~magnitude from V~magnitude would have additional uncertainties.}.
Some objects may have larger uncertainties on 
H~magnitudes if the MPC magnitudes are driven
by measurements from uncalibrated surveys, so
we assign a global uncertainty of 0.3~mag
on all H~magnitudes.
(This is likely an overestimate of the
uncertainties, though, since in most cases
a significant fraction of the reported photometry
comes from our W84 measurements, which have a 
relatively small uncertainty of around~0.1~mag.)

Another source of uncertainty concerns photometry of trailed sources.
In our survey --- as in most other NEO surveys ---
isophotal magnitudes are reported. For trailed
objects, this typically underreports the brightness.
In our survey this affects both the real and 
implanted objects, so although we are internally
consistent in terms of our debiasing, all fast-moving
objects may actually be somewhat (perhaps a few tenths
of a magnitude) brighter than reported. This error
bar is on the order of the largest error described
above.

Consequently, the derived size
distribution shown in Figure~\ref{sizedist} has an
uncertainty in the horizontal direction on the order
of 0.3~mag. In other words, our result should be written as
$10^{6.6}$~objects
larger than $H=27.3\pm0.3$, corresponding to
diameters of $10^{+2}_{-1}$~meters.

We have assumed that the average albedo in the NEO
population is~0.2.
This is the
average albedo 
for objects smaller than 200~m
as derived from NEOWISE observations
\citep{mainzer}.
Although the NEOWISE survey is relatively
insensitive to albedo, there is still a small
bias against high albedo objects.
Furthermore, the mean albedo for
very small NEOs could be different from~0.2; if 
these smallest bodies have fresh surfaces (either through
surface resetting due to planetary encounters, or through
being collisionally young objects) then the mean
albedo could be higher. Conservatively, we recalculate
the above steps using mean albedos of~0.4 and, for completeness,~0.1. 
The result is that $H=27.3$ corresponds to diameters
of 7--14~meters.
This uncertainty dominates that from the previous
paragraph.
We 
conservatively express our final
result as $10^{6.6}$~NEOs larger than
$10\pm4$~meters.

The uncertainties in our measured detection
efficiencies are small because we have many
thousands of objects for each night.
However, 
there is uncertainty in our cumulative
number of debiased NEOs.
There are 257~NEOs with $H\leq28$ in our
debiased sample. Formally, Poissonian statistics
implies a fidelity of around 6\% 
on this measurement. Conservatively, we assign
10\% uncertainties, which allows for other uncertainties
that may be present in our result. This
10\% uncertainty now also includes the possible
underestimate discussed at the end
of \S5.1.

%Our derived cumulative size distribution 
%(Fig.~\ref{sizedist})
%shows a ``flat'' section in the range
%19$<$H$<$21. This is probably caused by
%small number statistics: we did not detect
%many objects in this range (Fig.~\ref{hmag}).
%This same result can also be seen in Figure~\ref{HV}.
%There is no obvious bias in our survey
%against objects with these mid-range H values.
%However, to understand the effect of a
%possible un-characterized bias in this size
%range, we can artificially increase the 
%H=21 point in our derived size distribution
%by a factor of~2.5 (an extrapolation
%of the bright end slope)
%to eliminate this ``flat'' section.
%This corresponds to a factor of~2.5 for all sizes
%smaller than H=21. Thus, the resulting number
%of NEOs larger than 10~m could be underestimated
%by as much as a factor of~2.5.

Our final result from this analysis is therefore
that there are 
$10^{6.6}$$\pm$10\%~NEOs larger than
$10\pm4$~meters. We note that in the 
context of our survey this result is still preliminary,
as our 20~telescope nights from 2015 and 2016
will serve as independent measurements of
the size distribution and allow us to refine our
uncertainties, particularly in size regimes
where we presently have a relatively
small number of detections.

\subsection{Comparison to previous estimates}

\begin{figure}
\includegraphics[angle=270,scale=0.6]{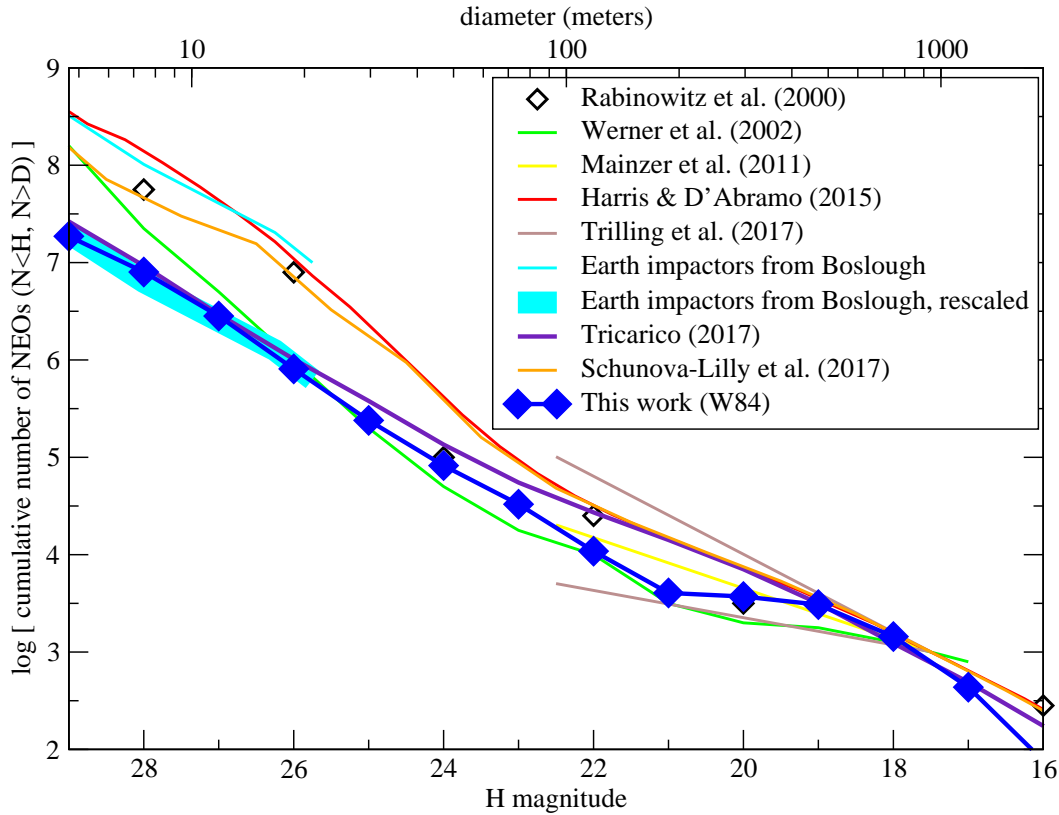}
\caption{
{\small The cumulative debiased size distribution of NEOs.
Our new result is shown as the thick blue line and data
points.
(Note that
while our results are shown
as binned data points, this binning is for
representational purposes only. Each object
is debiased individually according to its V~magnitude,
rate of motion, and distance, as described in the 
text.)
We find
around $10^{6.6}$~(or $3.5\times 10^6$) NEOs larger than
H=27.3 (around
10~meters), and
$10^{6.9}$~(or $7.9\times 10^6$) NEOs larger than H=28 (around
7~meters),
assuming an albedo
of~0.2 for each object. 
The error bars are $\sim$15\%
for H$>$20 and formally $\sim$5\% at H$>$28;
we conservatively take the overall error
bar at any given size to be 10\% (see text).
Our solution is normalized to have
1000~objects larger than 1~km
(H$\approx$17.5 for albedo of~0.2), which
is taken from the recent
NEOWISE \citep{mainzer}
and Spitzer/ExploreNEOs \citep{trilling} results.
Our result is in agreement with all
previous estimates for H$<$22,
and for H$>$24 is in particularly good agreement
with the recent result from \citet{tricarico}.
The \citet{harrisdabramo} and \citet{rabinowitz}
results have few data points for H$>$23.
The \citet{werner} result assumes a size-independent
impact probability that is equal to that for
the largest NEOs, but our new result implies
that the smallest NEOs may have a higher
impact probability, requiring fewer objects to
match the observed impactor population.
The thin light blue line shows impactors from
\citet{boslough} that have been normalized
using this same size-independent impact
probability. The slope of the \citet{boslough}
result agrees extremely well with our measured
size distribution, and the impact probability
for small NEOs can be re-derived by 
determining the normalization factor between
the \citet{boslough} bolide measurements and
our results in this size range. The thick light
blue line shows the \citet{boslough} result,
rescaled to our observations; the slopes agree
very well.
Our result implies a factor of ten fewer
Chelyabinsk-sized impactors than
previously estimated, though with a 
factor of ten greater impact probability than
previously assumed.
}
\label{sizedist}}
\end{figure}

There have been a number of previous estimates 
of the NEO size distribution, using a variety of 
techniques. We briefly describe previous
work here, and plot their independently 
derived size distributions in Figure~\ref{sizedist}.
All previous work agrees that there are
approximately 1000~NEOs larger than 1~km,
so we have normalized all the data described
in this section to have 1000~NEOs at H=17.5
(after \citealt{trilling}).

\citet{rabinowitz1993} made an
early estimate that was updated in
\citet{rabinowitz}, in both cases
using data from NEO surveys in a study
that is roughly analogous to ours, although
without implanting synthetic
objects into the data stream.
\citet{rabinowitz} present results down
to H=30, but there are very few objects
with H$>$24
included in their solution.
For H$<$24 our results agree with those
from \cite{rabinowitz} very well.

More recently, \citet{schunova}
have estimated the size distribution
based on Pan-STARRS1 data as debiased
through the (theoretical) population
model of \citet{greenstreet}. Their
population estimate is somewhat greater
than ours, though they do show a 
change to shallower slope (and therefore
relative deficit of NEOs compared
to \citealt{harrisdabramo} for 
H$>$24).

Both the Spitzer/ExploreNEOs and NEOWISE
teams have made independent measurements of
the size distribution of NEOs as a result of
their (independent) thermal infrared surveys.
NEOWISE results suggest
20,500$\pm$3000~NEOs larger than 100~m \citep{mainzer};
we find here around 18,000~NEOs larger than
100~m, in good agreement with the earlier result.
The nominal ExploreNEOs
result also suggests around 20,000~NEOs
larger than 100~m, with an acceptable
range of 5000--100,000 \citep{trilling};
we are again in close agreement here with 
the previous work.

\citet{harris2015},
\cite{boslough}, and \citet{harrisdabramo}
use
re-detection simulations
of ongoing ground-based surveys to estimate
completeness and therefore the
underlying NEO size distribution.
However, they have
no simulated observations (re-detections)
for objects with $H>25$, and the
completion rate is small for
$H>23$ \citep{boslough}.
Their estimate of the number
of 10~meter NEOs
is therefore
essentially an extrapolation
from larger
sizes. Our result agrees moderately
well with these re-detection results
for $H<23$, but the two results diverge
for $H>23$, where their number of
re-detections is small. This implies
that their extrapolation from larger sizes
may not be appropriate.

\citet{tricarico} used a similar
redetection approach and combined
two decades' worth of nine different
survey programs to estimate the underlying
NEO population. This result is seen
in Figure~\ref{sizedist} to be in extremely
close agreement with our result, especially
at the smallest sizes (H$>$26).

\citet{werner} analyzed the lunar crater
population to determine the size distribution
of the lunar impactor population, i.e.,
NEOs, as averaged over the past few billion
years. 
They estimate the NEO population down to 
10~meters, though with two significant
caveats: (1) the average impact probability
per asteroid is derived from the dynamical
calculations for the $\sim$1000~largest 
NEOs and applied to NEOs of all sizes, and
(2) 
there is some uncertainty whether
the current NEO population is the same as the 
historically averaged NEO population.
Our result agrees closely with \citet{werner}
for $H<26$. We do not reproduce their sharp
rise for 
$H>26$.

Finally, 
\citet{brownetal} analyzed impact data,
including both the Chelyabinsk impact and
various other data from the past several decades, to
deduce the size distribution
of Earth impactors (that is, very small NEOs).
A further analysis of this impactor
work is also presented in more detail
in \citet{boslough}.
This analysis produces a size distribution of
impactors covering the approximate range 1--20~meters.
To convert from impactors
to the entire NEO population, a
scaling factor that has its origins in the calculated
annualized impact rate of the 1000~largest
NEOs is used (as \citealt{werner} did).
The factor used to date places the impactor
size distribution in agreement with the \citet{harris2015}
and \citet{harrisdabramo}
NEO size distribution extrapolation in that size range.

Our measured slope from 1--20~meters is 
identical to that from \citet{brownetal}
and \citet{boslough},
but there is an offset in the absolute number:
we find $10^{6.9}$~NEOs with
$H>28$ compared to their
$10^{8}$.
Their result, using their scaling 
from the largest NEOs,
is shown as the thin cyan
line in Figure~\ref{sizedist}.
The thick cyan line in Figure~\ref{sizedist}
shows their impactor data normalized
to our measured size distribution at H=26 (around
20~meters).
The slopes of their impactor size distribution
and our measured NEO size distribution agree
extraordinarily well.

\subsection{Implications of our result}

This is the first time that a 
single observational data set has been 
used to measure the size distribution
from 1~km down to 10~meters. Previous 
direct measurement work
had data in either the larger ($>$300~m) or,
indirectly, the
small ($<$20~m) regime. 
Very broadly, our result is in agreement
with most of the previous work, but there
are several aspects that warrant further
exploration.

The number of $\sim$10~meter-sized
NEOs is of keen interest because these
objects impact the Earth relatively frequently
and can cause severe damage (as happened
in Chelyabinsk, Russia, in 2013).
At this size range
our results appear to disagree with those
of \citet{rabinowitz} and \citet{harrisdabramo},
but both of these have very few data points
for H$>$23.
%The primary significant difference between
%our new result and previous results is for
%H$>$25. 
For H$<$25, our result agrees very
well with the \citet{werner} result, but 
they report a strong upturn at H=26 that is 
not seen in \citet{tricarico}, and \citet{schunova}
show a relative downturn at that same size.

The \citet{brownetal} and \citet{boslough}
results for impactors in the size range 1--20~m
are significantly higher than our derived
result. 
\citet{werner}, \citet{brownetal}, and \citet{boslough}
all have in common
the assumption that the impact rate of the 
smallest NEOs is the same as that of
the largest NEOs. If instead the impact
rate of the smallest NEOs is an order of
magnitude greater than that of the biggest
NEOs, then the number of small NEOs implied
% in
%\citet{werner}, \citet{brownetal}, and
%\citet{boslough} 
would be reduced by that same
factor. The impact rate of small NEOs
could be larger than their large counterparts
if the orbit distributions of those two
populations differ, for 
example if
there exist bands of collisional debris
or meteoroids in orbits similar to that 
of the Earth that the Earth spends
significant time transiting,
similar to meteor streams but with a different origin
(A.\ Harris [DLR], pers.\ comm.).
This could result, for example, from the fragmentation
of medium-small NEOs into swarms of smaller
boulders that pass near the Earth, such as
is implied by results reported in
\citet{mommertBD,mommertMD}.
A very recent result \citep{spurny} suggest 
that one such band in which there is 
a relative enhancement of $\sim$10~meter-sized
NEOs
indeed may exist, and
\citet{jeongahn} find that the lunar
cratering rate is higher for 1--10~meter NEOs
than for kilometer size objects.
The sharp upturn in the
\citet{werner} distribution at H=26 
(Figure~\ref{sizedist}) might 
reflect an increasing impact probability at 
that size.
We emphasize that the slope of the 
\citet{brownetal} and \citet{boslough}
results for bodies 1--20~km matches
our derived slope very well, implying that
the discrepancy arises not from measurements
of the size distribution but the normalization
assumption used in their work.

We note that our data point at H=21 appears
depressed compared to the adjacent points and
the overall implied continuum slope. This 
dip has been seen in other work as well
\citep{werner,harrisdabramo,schunova}.
\citet{harrisdabramo} offer two plausible
explanations. The first is that this size
(around 100--200~meters) corresponds to
the transition from weak rubble pile
asteroids at larger sizes to stronger 
monolithic bodies at small sizes, and that
the relative deficit of bodies at this size
indicates the maximally disrupted asteroid size.
Their second proposed explanation is that
if there is a shift in average albedo at this
size (perhaps due to collisions among smaller
but not larger bodies) then the conversion
between H~magnitude and diameter would naturally
produce an apparent dip.

%Finally, we can update the impact risk,
%based on our new results.
%We find that the 
%impact risk of
%objects in the 10--20~m range --- the size
%of the Chelyabinsk impactor --- is approximately
%ten times less than has been previously
%estimated. 
%We scale the \citet{boslough} result to find
%a new estimate that 
%Chelyabinsk-sized
%impactors hit the Earth on average
%every $\sim$300~years.

\section{Conclusions and future work}

We are carrying out a 30~night survey to 
detect NEOs with the Dark Energy Camera
and the 4-meter Blanco telescope at CTIO.
In year~1 we made 1377~measurements of 
235~unique NEOs. Through implanting synthetic
objects in our data stream and measuring the
detection efficiency of our survey as a function
of magnitude and rate of motion, we have 
debiased our survey. 
We find that there are 
around $10^{6.6}$~(or $3.5\times 10^6$) NEOs larger than
H=27.3 (around
10~meters), and
$10^{6.9}$~NEOs larger than H=28 (around
7~meters). 
This population estimate is around a 
factor of ten less than has been previously
estimated, though in close agreement with one
recent measurement and
somewhat in agreement with another.
Our derived NEO size distribution
--- the first to cover the entire range from
1~km to 10~meters based on a single
observational data set ---
matches basically all observed
data for sizes larger than 100~meters.
In the size range 1--20~m, our measured
slope matches the bolide impactor slope
quite closely, and implies that the impact
probability for any given small NEOs is greater
than that for a large NEO by a factor
of ten or more. 
%The newly calculated impact
%risk for Chelyabinsk-sized objects is a factor
%of ten less than estimated in \cite{boslough} and
%other work.

We have data from 10~survey nights in each of 2015 and 2016
that are not
analyzed here, though all observations
have already been reported 
to the MPC. These more recent data will
be used to independently measure the size distribution
and refine the error bars on the estimate presented here.
We will also extend our analysis to W84
objects that were detected on only one or two
nights and are therefore not designated by the MPC.
This overall experiment is in some ways a pathfinder for
the upcoming Large Synoptic Survey Telescope (LSST),
which has NEO observations as one of its
primary science drivers. LSST will have a much bigger
aperture than the Blanco (8.4~meters compared
to 4~meters), and cover far more sky than we have 
here (20,000~deg$^2$ compared to our $\sim$975~deg$^2$), making it the most 
comprehensive NEO survey ever carried out. When software
tools capable of implanting and detecting synthetic
objects are in place, a very high fidelity 
measurement of the NEO size distribution to sizes
as small as 1~meter will be possible.


\acknowledgments

We thank Peter Brown and Alan Harris (DLR) for many useful conversations
and Steve Chelsey and an anonymous AAS statistics
reviewer for useful comments that improved this paper.
We thank the NOAO TAC and Director
Dave Silva for granting Survey status for this program.
We also thank Dave Silva and NOAO for acquiring the VR filter that we 
use in our survey.
We gratefully acknowledge the hard work and 
help that Tim Spahr and Gareth Williams of the
Minor Planet Center have provided and continue
to provide in support of
our DECam NEO survey.
DET carried out some of the work on this paper
while being hosted at Lowell Observatory.
This work was supported
in part by NASA award NNX12AG13G.
This work is based
on observations at Cerro Tololo Inter-American Observatory, National Optical Astronomy Observatory (NOAO Prop.\ 2013B-0536; PI: L.\ Allen), which is operated by the Association of Universities for Research in Astronomy (AURA) under a cooperative agreement with the National Science Foundation.

This project used data obtained with the Dark Energy Camera (DECam), which was constructed by the Dark Energy Survey (DES) collaboration.
Funding for the DES Projects has been provided by 
the U.S. Department of Energy, 
the U.S. National Science Foundation, 
the Ministry of Science and Education of Spain, 
the Science and Technology Facilities Council of the United Kingdom, 
the Higher Education Funding Council for England, 
the National Center for Supercomputing Applications at the University of Illinois at Urbana-Champaign, 
the Kavli Institute of Cosmological Physics at the University of Chicago, 
the Center for Cosmology and Astro-Particle Physics at the Ohio State University, 
the Mitchell Institute for Fundamental Physics and Astronomy at Texas A\&M University, 
Financiadora de Estudos e Projetos, Funda{\c c}{\~a}o Carlos Chagas Filho de Amparo {\`a} Pesquisa do Estado do Rio de Janeiro, 
Conselho Nacional de Desenvolvimento Cient{\'i}fico e Tecnol{\'o}gico and the Minist{\'e}rio da Ci{\^e}ncia, Tecnologia e Inovac{\~a}o, 
the Deutsche Forschungsgemeinschaft, 
and the Collaborating Institutions in the Dark Energy Survey. 
The Collaborating Institutions are 
Argonne National Laboratory, 
the University of California at Santa Cruz, 
the University of Cambridge, 
Centro de Investigaciones En{\'e}rgeticas, Medioambientales y Tecnol{\'o}gicas-Madrid, 
the University of Chicago, 
University College London, 
the DES-Brazil Consortium, 
the University of Edinburgh, 
the Eidgen{\"o}ssische Technische Hoch\-schule (ETH) Z{\"u}rich, 
Fermi National Accelerator Laboratory, 
the University of Illinois at Urbana-Champaign, 
the Institut de Ci{\`e}ncies de l'Espai (IEEC/CSIC), 
the Institut de F{\'i}sica d'Altes Energies, 
Lawrence Berkeley National Laboratory, 
the Ludwig-Maximilians Universit{\"a}t M{\"u}nchen and the associated Excellence Cluster Universe, 
the University of Michigan, 
{the} National Optical Astronomy Observatory, 
the University of Nottingham, 
the Ohio State University, 
the University of Pennsylvania, 
the University of Portsmouth, 
SLAC National Accelerator Laboratory, 
Stanford University, 
the University of Sussex, 
and Texas A\&M University.


\vspace{5mm}
\facility{Blanco(DECam)}


\software{NOAO DECam Community Pipeline (CP) \citep{valdes} with Moving Object Detection System (MODS) \citep{mods}}

\clearpage

%% the file, uncomment the \plotone, \plottwo, and
%% If you need a layout that cannot be achieved with \plotone or
%% \plottwo, you can invoke the graphicx package directly with the
%% \includegraphics command or use \plotfiddle. For more information,
%% please see the tutorial on "Using Electronic Art with AASTeX" in the
%% documentation section at the AASTeX Web site,
%% http://www.journals.uchicago.edu/AAS/AASTeX.
%%
%% The examples below also include sample markup for submission of
%% supplemental electronic materials. As always, be sure to check
%% the instructions to authors for the journal you are submitting to
%% for specific submissions guidelines as they vary from

%% This example uses \plotone to include an EPS file scaled to
%% 80% of its natural size with \epsscale. Its caption
%% has been written to indicate that additional figure parts will be
%% available in the electronic journal.

\clearpage

\clearpage

\clearpage

\clearpage

\clearpage

\clearpage

\clearpage

\clearpage

\clearpage

\end{document}